\documentclass[aps,prb,onecolumn,nofootinbib,notitlepage,superscriptaddress,longbibliography,preprintnumbers]{revtex4-2}
\usepackage{times}
\usepackage{graphicx}
\usepackage[whole]{bxcjkjatype} 

\usepackage[top=30truemm,bottom=30truemm,left=25truemm,right=25truemm]{geometry}

\usepackage[colorlinks=true,linktoc=page,citecolor=red,linkcolor=blue]{hyperref}	
\graphicspath{{./figs/}}  
\usepackage[usenames]{xcolor}
\usepackage{amsmath,amssymb,amsfonts}	
\usepackage{bm,braket,array}
\usepackage{multirow}
\usepackage[all]{xy}
\usepackage{amsthm}
\usepackage{amscd}
\usepackage{slashed} 
\usepackage{calligra} 
\usepackage[normalem]{ulem} 

\theoremstyle{definition}

\theoremstyle{remark}

\def\pf{{\rm Pf\,}}
\def\mod{{\rm\ mod\ }}

\newcommand{\G}{\Gamma}

\newcommand{\R}{\mathbb{R}}
\newcommand{\Z}{\mathbb{Z}}

\newcommand{\bk}{{\bm{k}}}

\def\widebar{\accentset{{\cc@style\underline{\mskip10mu}}}} 
\def\wideubar{\underaccent{{\cc@style\underline{\mskip10mu}}}} 

\begin{document}
\title{A discrete formulation of the Kane-Mele $\Z_2$ invariant}
\author{Ken Shiozaki}
\affiliation{Center for Gravitational Physics and Quantum Information, Yukawa Institute for Theoretical Physics, Kyoto University, Kyoto 606-8502, Japan}
\date{\today}
\preprint{YITP-23-61}

\begin{abstract}
We present a discrete formulation of the Kane-Mele $\Z_2$ invariant that is manifestly gauge-independent and quantized.
\end{abstract}
\maketitle
\parskip=\baselineskip

In this short note, we present a formulation of the Kane-Mele $\Z_2$ invariant~\cite{KaneMeleZ2} that does not require any gauge fixing conditions and is quantized for any discrete approximation of the Brillouin zone (BZ) torus. 
A known gauge-fixing-free method involves tracking the flow of the center positions of the hybrid Wannier states or extracting one eigenvalue from each Kramers pair among the eigenvalues of the Wilson loops~\cite{Soluyanov_11, Yu_11}.
The construction below is based on the time-reversal polarization~\cite{Fu_Kane_TR_polarization_06} and eliminates the time-reversal gauge condition in \cite{Fukui_Hatsugai_Z2}. 

We denote the Bloch momentum in the BZ torus by $\bk=(k_x,k_y) \in [-\pi,\pi] \times [-\pi,\pi]$.
Consider a $2N \times 2N$ Hamiltonian $H(\bk)$ that satisfies time-reversal symmetry (TRS).
The time-reversal operator is defined as $T=U_TK$, where $U_T$ is a $2N \times 2N$ unitary matrix that satisfies $(U_T)^{\rm tr} = -U_T$, and $K$ represents complex conjugation.
TRS for $H(\bk)$ is expressed as $TH(\bk)T^{-1}= H(-\bk)$.
Due to TRS, the independent region of BZ is the cylinder ${\cal C} = \{(k_x,k_y)|k_x \in [-\pi,\pi], k_y \in [0,\pi]\}$.
We introduce a lattice approximation $|{\cal C}|$ of the cylinder ${\cal C}$, with $\delta k_x$ and $\delta k_y$ as the lattice spacings in the $k_x$ and $k_y$ directions, respectively, ensuring that $|{\cal C}|$ includes the high-symmetry points $\bk = (0,0), (\pi,0), (0,\pi), (\pi,\pi)$.
Denote the energy eigenvalues of the Hamiltonian $H(\bk)$ at Bloch momentum $\bk$ as $E_1(\bk) \leq \cdots \leq E_{2N}(\bk)$ in ascending order. 
We focus on a set of $2n$ isolated bands $2p+1, \dots, 2p+2n$, such that for any $\bk \in [-\pi,\pi] \times [-\pi,\pi]$, the inequality $E_{2p}(\bk) + \Delta< E_{2p+1}(\bk) \leq \cdots \leq E_{2p+2n}(\bk) < E_{2p+2n+1}(\bk)-\Delta $ holds with $\Delta>0$ a finite energy gap.
At the lattice points $\bk$, compute the orthonormal set of $2n$ eigenstates ${\cal U}(\bk)= \big(\ket{u_{2p+1}(\bk)}, \dots,\ket{u_{2p+2n}(\bk}\big)$ of $H(\bk)$, with $H(\bk)\ket{u_j(\bk)} = E_j(\bk) \ket{u_j(\bk)}$ for $j=2p+1,\dots,2p+2n$. 
In the following, we define the $\Z_2$ invariant $\nu \in \{0,1\}$ that is invariant under the $U(2n)$ gauge transformations 
\begin{align}
    {\cal U}(\bk) \mapsto {\cal U}(\bk) V(\bk),\quad V(\bk) \in U(2n).
    \label{eq:gauge_tr}
\end{align}
We introduce the unitary matrix
\begin{align}
    w(\bk)= {\cal U}(-\bk)U_T {\cal U}(\bk)^* \in U(2n). 
    \label{eq:w}
\end{align}
Note that $w(\bk)^T=-w(-\bk)$, and thus at the high-symmetric points $\bm{\G} \equiv -\bm{\G}$, the Pfaffian $\pf[w(\bm{\G})] \in U(1)$ is defined.
Furthermore, at the high-symmetry points $\bm{\G}$, the gauge transformation (\ref{eq:gauge_tr}) changes $\pf[w(\bm{\G})]$ as $\pf [w(\bm{\G})] \mapsto \pf[w(\bm{\G})] \det [V(\bm{\G})]^*$.

At the boundary discrete circles of the cylinder $|{\cal C}|$, namely the lattice points on the $k_y=0$ and $k_y=\pi$ lines, the time-reversal polarization is defined~\cite{Fu_Kane_TR_polarization_06}:
\begin{align}
    P_{T,x}(\G_y)=\frac{1}{2\pi} {\rm arg} \left[ \prod_{k_x=0}^{\pi-\delta k_x} \det[{\cal U}(k_x+\delta k_x,\G_y)^\dag {\cal U}(k_x,\G_y)] \frac{\pf[w(0,\G_y)]}{\pf[w(\pi,\G_y)]} \right] \in \R/\Z,\quad \G_y = 0,\pi.
\end{align}
Here, ${\rm arg}(z)$ is the argument of the complex number $z$.
Note that the time-reversal polarization $P_{T,x}(\G_y)$ does not depend on the gauge choice of ${\cal U}(\bk)$ and is well-defined modulo 1.
Using ${\cal U}(-\bk)=U_T {\cal U}(\bk)^* w(\bk)^\dag$, we find that the usual polarization
\begin{align}
    P_x(\G_y)=\frac{1}{2\pi}{\rm arg} \left[\prod_{k_x=-\pi}^{\pi-\delta k_x} \det[{\cal U}(k_x+\delta k_x,\G_y)^\dag {\cal U}(k_x,\G_y)] \right] \in \R/\Z,\quad \G_y = 0,\pi, 
\end{align}
has the following relation to the time-reversal polarization $P_{T,x}(\G_y)$: 
\begin{align}
P_x(\G_y) = 2 P_{T,x}(\G_y) \mod 1.
\label{eq:gt_g}
\end{align}
Define the Berry flux through the plaquette $\Box_\bk$ formed by the four lattice points $(k_x,k_y), (k_x+\delta k_x, k_y),(k_x+\delta k_x,k_y+\delta k_y), (k_x,k_y+\delta k_y)$ as
\begin{align}
F(\Box_\bk) 
&= 
{\rm arg}\Big[
\det[{\cal U}(k_x,k_y)^\dag {\cal U}(k_x,k_y+\delta k_y)]
\det[{\cal U}(k_x,k_y+\delta k_y)^\dag {\cal U}(k_x+\delta k_x,k_y+\delta k_y)]\nonumber\\
&\det[{\cal U}(k_x+\delta k_x,k_y+\delta k_y)^\dag {\cal U}(k_x+\delta k_x,k_y)]
\det[{\cal U}(k_x+\delta k_x,k_y)^\dag {\cal U}(k_x,k_y)] \Big] \in \R. 
\end{align}
The Berry flux $F(\Box_\bk)$ is gauge-invariant. 
Moreover, even though $F(\Box_\bk)$ is originally an $\R/2\pi \Z$-valued quantity, if the plaquette $\Box_\bk$ is sufficiently small compared to the variation of the Hamiltonian $H(\bk)$, $F(\Box_\bk)$ is close to $0$ as an $\R/2\pi \Z$ value and can be regarded as a real number.
The sum of the Berry fluxes over the plaquettes $\Box_\bk$ in the cylinder $|{\cal C}|$ exactly matches the difference in polarization $P(\G_y)$ at the boundaries of the cylinder:
\begin{align}
    \frac{1}{2\pi} \sum_{\Box_\bk \in |{\cal C}|} F(\Box_\bk) \equiv P_x(0)-P_x(\pi) \quad \mod 1.
\end{align}
From (\ref{eq:gt_g}), the $\Z_2$ invariant $\nu \in \{0,1\}$ is given by
\begin{align}
    \nu= \frac{1}{2\pi}\sum_{\Box_\bk \in |{\cal C}|} F(\Box_\bk) - 2 P_{T,x}(0)+ 2P_{T,x}(\pi) \in \{0,1\} \quad \mod 2
\end{align}
The $\Z_2$ invariant $\nu$ does not depend on the gauge of ${\cal U}(\bk)$, and any time-reversal symmetric gauge condition is not necessary.
Moreover, by construction, the $\Z_2$ invariant $\nu$ is quantized for any discrete approximation of the BZ torus.

\bibliography{ref}

\end{document}